\documentclass[twocolumn,showpacs,preprintnumbers,amsmath,amssymb]{revtex4}
\usepackage{graphicx}
\usepackage{dcolumn}
\usepackage{bm}
\begin{document}

\title{Spin precession in a fractional quantum Hall state with
spin-orbit coupling}
\author{Marco Califano$^{\ast}$, Tapash Chakraborty$^{\ast\ddag}$
and Pekka Pietil\"ainen$^{\ast\dag}$}
\affiliation{$^\ast$Department of Physics and Astronomy,
University of Manitoba, Winnipeg, Canada R3T 2N2}
\affiliation{$^\dag$Department of Physical Sciences/Theoretical Physics,
P.O. Box 3000, FIN-90014 University of Oulu, Finland}
\date{\today}

\begin{abstract}
Experimental attempts to realize spin-devices based on concepts
derived from single-particle theoretical approaches have not been
very successful yet. This raises the fundamental question of whether
inter-electron interactions can be neglected in planar electron-based
spintronics devices. We report on our results of a many-body approach
to the spin configuration in a quantum Hall state in the presence of
Bychkov-Rashba type spin-orbit interaction. While some properties of this
system are found to be ideally suited for exploitation in spin devices,
others might seem to limit its applicability. The latter can
however be optimized for device performance.
\end{abstract}
\maketitle

Spintronics has become a fast developing field in which the electron
spin degree of freedom is exploited to create novel electronic devices
\cite{awschalom_book}. Of paramount importance in this pursuit is the
ability to manipulate spins in a controlled and reliable way. The
spin-orbit (SO) interaction provides a bridge between spin and charge
properties which, by coupling the electron momentum with its spin, makes
it possible to control the spin dynamics using electric fields. After
the initial proposal of a spin device based on this concept
\cite{datta_apl56}, the scientific literature has been saturated by
theoretical studies on the effects of SO coupling in inversion-asymmetric
two-dimensional systems. Most of these \cite{winkler_prb69,koga_prbR70},
however, rely on a simplified single-particle picture, where
electron-electron interaction is neglected. Although many different
devices based upon this paradigm have been proposed theoretically
the experimental realization of much of them has remained elusive. This
raises a question about the wisdom of ignoring the influence of many-body
effects on the spin configuration.

A crucial factor that has also contributed to such a delay in the
realization of many theoretical schemes is the short spin lifetime in
conventional two-dimensional semiconductors. A 2DEG in the Quantum
Hall (QH) state with odd integer ($\nu=1$) or fractional ($\nu=1/3$)
filling factor, however, is completely spin-polarized \cite{book} and there
is no spin scattering in such a system. The mobility is very high and
therefore there is also very little scattering from impurities. This
removes the problem of spin decoherence found in conventional 2DEG
systems due to spin decay mechanisms such as the Dyakonov-Perel relaxation
\cite{dyakonov_jetp33}. For the same reason (i.e., the spin polarization),
these QH states might well be efficient spin injectors.

The special features of Quantum Hall (QH) systems with integer ($\nu=1$)
filling factors have already found application in different spin devices:
the use of edge states in the QH regime has been proposed as spin polarizer
for spin read-out in single-spin memory devices \cite{recher_PRL85}, to
achieve pure-state initialization of the qubit \cite{vandersypen_quantph},
to enhance Coulomb Blockade measurements \cite{ciorga_condmat9912446},
and more generally as an efficient spin injector into semiconductors
\cite{murakami_science301}. Furthermore, weak spin relaxation in QH states
was exploited recently in the design of spin devices by Pala et al.
\cite{pala_PRB71}. All these applications exploit the absence (or weakness)
of spin relaxation mechanisms in QH systems and the high spin-polarization
of QH states. Despite increasing interest in QH systems for spintronics
applications and the important role played in this context by the SO
interaction in a 2DEG, no many-body study on the spin configuration in QH
states in the presence of SO coupling exists to date.

In this letter we present a theoretical investigation into the effects
of Bychkov-Rashba SO coupling in the presence of Coulomb interactions
on the spin configuration in a fractional quantum Hall (FQH) state.
Our approach is a generalization of the spin precession concepts
developed in a recent paper by Koga and co-workers \cite{koga_prbR70}
to a fractional quantum Hall (FQH) system \cite{book,noi_cond-mat}.
According to that simple single-particle picture of Koga et al.,
assuming the electron wave vector {\bf k}$||{\hat{x}}$, and
taking the spin basis along the $z$ axis perpendicular to the 2DEG
plane, the two (Bychkov-Rashba) spin split states with energies
$E=\hbar^2k^2/2m^* \pm \alpha k$, can be written as
\begin{eqnarray}
{\bf \Psi}_{{\bf k}\uparrow} &=& \frac{1}{2} \left(\begin{array}{c}1-i\\1+i
\end{array}\right)e^{ikx}\label{eq:spin1}\\
{\bf \Psi}_{{\bf k}\downarrow} &=& \frac{1}{2} \left(\begin{array}{c}1
+i\\1-i\end{array}\right)e^{ikx}.\\
\nonumber
\label{eq:spin2}
\end{eqnarray}
One can then build a linear combination of these two states at a given
energy (in this case the wave vectors are written as ${\bf k}=(k\pm\Delta
k,0,0)$, respectively, for spin-down and -up states):
\begin{eqnarray}
{\bf \Psi}_{\bf k} &=& \frac{1}{\sqrt{2}}({\bf \Psi}_{{\bf k-\Delta k}
\uparrow}+ {\bf \Psi}_{{\bf k+\Delta k}\downarrow})\label{eq:lincomb}\\
&=&\frac{1}{2\sqrt{2}} \left\{\left(\begin{array}{c}1-i\\1+i\end{array}
\right)e^{i(k-\Delta k)x}+\left(\begin{array}{c}1+i\\1-i\end{array}\right)
e^{i(k+\Delta k)x}\right\}\nonumber\\
&=&e^{ikx}\frac{1}{\sqrt{2}}\left(\begin{array}{c}\cos(\Delta k\,x)-
\sin(\Delta k\,x)\\\cos(\Delta k\,x)+\sin(\Delta
k\,x)\end{array}\right)\label{eq:precession}
\end{eqnarray}
whose spin orientation depends on the position along $x$ and on the
strength of the spin-orbit interaction ($\Delta k\sim \alpha$).
Therefore, as ${\bf \Psi}_{\bf k}$ propagates along $x$ we expect the
spin to precess.

We have performed a many-body analog to this simple single-particle
analytical treatment of the effects of Bychkov-Rashba coupling in a
$\nu=1/3$ FQH state, where the many-body Schr\"odinger equation was
solved by means of the exact diagonalization scheme for four electrons
per supercell \cite{book}. The many-body wavefunctions were expanded in
terms of a complete basis obtained as superposition of solutions of the
single-particle Hamiltonian
\begin{equation}
H = \frac{({\bf p}-e{\bf A})^2}{2m^*}+\frac{\alpha}{\hbar}
[\mbox{\boldmath $\sigma$}\times({\bf
p}-e{\bf A})]_z+\frac{1}{2}g\mu_BB\sigma_z
\end{equation}
that includes the Bychkov-Rashba term \cite{rashba_jpc17}
and the Zeeman term. Here ${\bf p}$ is the momentum operator,
$\alpha$ is the SO coupling strength and
$\mbox{\boldmath $\sigma$}=(\sigma_x, \sigma_y, \sigma_z)$
are the Pauli spin matrices. Solutions of this Hamiltonian are also
spinors but have a more complex form than Eq. (\ref{eq:spin1}) and
Eq.~(\ref{eq:spin2}), due to the presence of the external magnetic field
(see \cite{noi_cond-mat} for details)
\begin{eqnarray*}
\psi_{s,j}^+({\bf r}) &=&
\frac1{\sqrt{\sqrt{\pi}l_0L_yA_s}}\sum_{n}\exp\left[i(X_j+nL_x)
\frac{y}{l_0^2}\right. \\
&-&\left.\frac{(X_j+nL_x-x)^2}{2l_0^2}\right] \\
&&\times\left(\begin{array}{c}
-iD_s\beta_{s-1}H_{s-1}\left(\frac{(X_j+nL_x-x)}{l_0}\right)\\
\beta_{s}H_{s}\left(\frac{(X_j+nL_x-x)}{l_0}\right)
\end{array}\right) \\
\psi_{s,j}^-({\bf r}) &=&
\frac1{\sqrt{\sqrt{\pi}l_0L_yA_s}}\sum_{n}\exp\left[i(X_j+nL_x)
\frac{y}{l_0^2}\right. \\
&-&\left.\frac{(X_j+nL_x-x)^2}{2l_0^2}\right] \\
&&\times \left(\begin{array}{c}
\beta_{s-1}H_{s-1}\left(\frac{(X_j+nL_x-x)}{l_0}\right)\\
-iD_s\beta_{s}H_{s}\left(\frac{(X_j+nL_x-x)}{l_0}\right)
\end{array}\right),
\end{eqnarray*}
and the corresponding energies
\begin{equation}
E_s^{\pm}=s\hbar\omega_c\pm\sqrt{E_0^2+2s\alpha^2/l_0^2}.\label{eq:Ekin}
\end{equation}
Here
\begin{equation}
D_s=\frac{\sqrt{2s}\alpha/l_0}{E_0+\sqrt{E_0^2+2s\alpha^2/l_0^2}}\label{eq:D}
\end{equation}
$A_s=1+D_s^2$, and $E_0=1/2(\hbar\omega_c-g\mu_BB)$,
$H_n(x)$ is the Hermite polynomial of degree $n$,
$l_0=(\hbar/m^*\omega_c)^{1/2}$ is the radius of the cyclotron orbit
with frequency $\omega_c=eB/m^*$ and center $X_j=k_yl_0^2$,
$L_xL_y\propto l_0^2$ is the supercell area, and $\beta_n=1/\sqrt{2^nn!}$.

\begin{figure}[h]
\begin{center}
 \includegraphics[width=.4\textwidth]{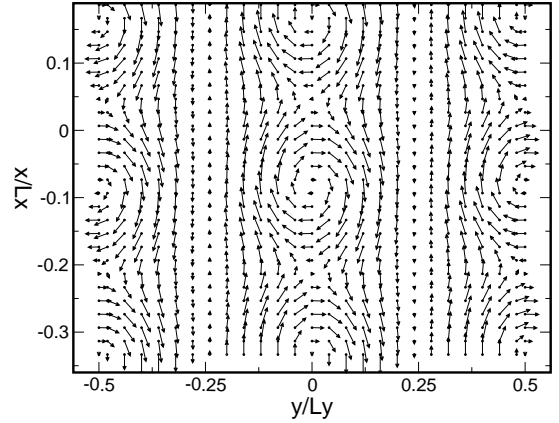}
\protect\caption{
Projection of $\langle{\bf S}\rangle$ on the $xy$ plane, calculated for
$\alpha=40$. Only a portion of the supercell is shown, as the behavior is
periodic in both directions (the motion is along the
$y$ direction).
}\label{fig:spins_xy}
\end{center}
\end{figure}

We then constructed a state as linear combination of two many-body wavefunctions,
eigenstates of Coulomb interaction, relative to two degenerate excited states
with different total momentum J (in analogy to Eq. (\ref{eq:lincomb}); in our case
however the two spins are not pointing in opposite directions due to the
spin polarized nature of the FQH state \cite{book,laughlin}),
and calculated the expectation value of the spin components
$S_x=\langle\sigma_x\rangle$, $S_y=\langle\sigma_y\rangle$ and $S_z=\langle
\sigma_z\rangle$
along the three principal directions.

\begin{figure}[h]
\begin{center}
 \includegraphics[width=.4\textwidth]{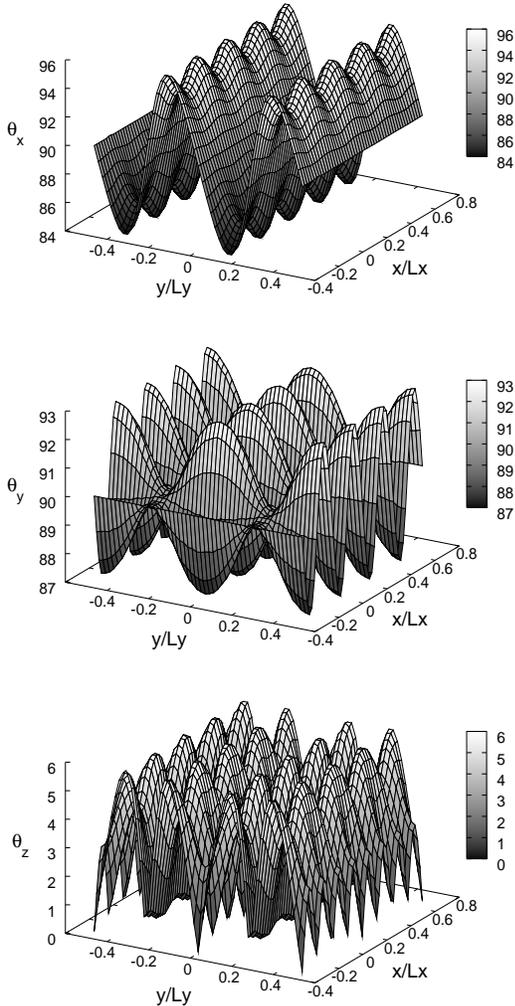}
\protect\caption{
Variation across the supercell of the angles between $\langle{\bf S}\rangle$
and the three principal axes, for $\alpha=40$.
}\label{fig:teta_3D}
\end{center}
\end{figure}

\begin{figure}[h]
\begin{center}
 \includegraphics[width=.35\textwidth]{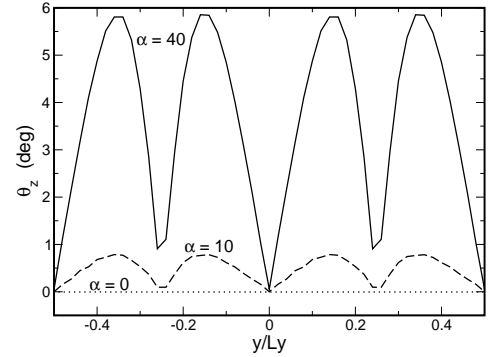}
\protect\caption{
Angle between $\langle{\bf S}\rangle$ and the $z$ axis as a function of
the position along the $y$ axis, for $\alpha=0$ (dotted line), 10 (dashed line),
40 nm$\cdot$meV (solid line) and $B=1$ T.
}\label{fig:teta}
\end{center}
\end{figure}

Our results (Fig.~\ref{fig:spins_xy}-\ref{fig:teta})
show a more complex picture than that described by Eq.~\ref{eq:precession}
for the resulting superposition of spin eigenstates, that, however, retains
the main feature of a position-dependent spin orientation  (i.e., the
precession). As shown in Fig.~\ref{fig:spins_xy} the spin rotation has
a period of $L_y/2$. Our calculated spin precession length is therefore
$L_{sp}= 4.1l_0$, which for $B=1$ T is of the order of 100 nm.
While $L_{sp}$ depends only on the applied {\em magnetic} field,
we find that the value of the different angles $\theta_i$
($i=x,y,z$), which the electron spin forms with the principal axes,
depends also on the applied {\em electric} field through the SO
coupling strength $\alpha$.

This property of the $\nu=1/3$ FQH state could be exploited in a spin
device (e.g. in a spin transistor). The use of such a state in a spin
device would, in fact, have two main advantages over that of conventional
2DEG systems: (i) no need for spin injection; (ii) no uncontrolled spin
decoherence (scattering) effects. (i) Although efficient spin injection
is fundamental for spintronic devices, it has been elusive so far
\cite{wolf_science294}. The $\nu=1/3$ FQH state, being {\em naturally}
spin-polarized would remove the need for spin injection altogether. In
fact, in the FQH state complete spin polarization is achieved via electron
correlations {\em without} any assistance from the Zeeman term
\cite{book,laughlin}. The initial phase of the spin (usually fixed at
the interface with the injector) is zero (i.e., the spin is parallel to
$B$) at the channel edge. This can be understood from Fig.~\ref{fig:spins_xy}
by noticing that for $y/L_y\sim \pm 0.25$ the spin in-plane component is
zero and, unlike anywhere else in the supercell, is the same for all
values of $x$. The position at $y/L_y\sim \pm 0.25$ is
due to the periodic boudary conditions used in the calculations.
(ii) As is the case for QH states with (odd) integer filling factors,
there is no spin scattering through the $\nu=1/3$ state, resulting
in a long spin lifetime.

In InAs structures, together with low spin scattering rates in the FQH state
(yet to be observed in conventional InAs 2DEGs), there is the added advantage
of a high {\em g}-factor. The only source of spin ``decoherence" in such a
FQH system is therefore introduced only by the electric-field-driven
Bychkov-Rashba field, that leads to a (position dependent) spin precession.
It follows that if the (ferromagnetic) drain contact has a spin polarization
${\bf P}_D$, an electron will be able to leave the channel (FQH state) only
if its spin at the end of the channel $\langle{\bf S}\rangle$ is aligned with
${\bf P}_D$. In our system we found that
\begin{equation}
\langle{\bf S}\rangle =
\left(\begin{array}{c}S_x(x,y,j,\alpha)\\S_y(x,y,j,\alpha)\\S_z(x,y,j,\alpha)
\end{array}\right),\label{eq:S}
\end{equation}
\noindent where $x$ and $y$ are the coordinates in the 2DEG plane,
$j$ is related to $k$ via $k_y=2\pi j/L_y$, $L_y$ is the supercell size
along $y$ and $\alpha$ is the Bychkov-Rashba coupling strength (the variation
of the angles that $\langle{\bf S}\rangle$ forms with the different axes,
as it precesses across the supercell, is shown in Fig.~\ref{fig:teta_3D}).
By choosing, for example, ${\bf P}_D || z$ we found that the angle between
$\langle{\bf S}\rangle$ and ${\bf P}_D$ is
\begin{equation}
\theta_z = \arccos(S_z(x,y,j,\alpha)/|\langle{\bf S}\rangle|).
\end{equation}
This angle is plotted in Fig.~\ref{fig:teta} for three values of $\alpha$.

A tunable device is obtained by varying the value of the Bychkov-Rashba
coupling constant $\alpha$, via the applied electric
field. As shown in Fig.~\ref{fig:teta}, this will induce a variation in
$\theta_z$: its value will change from 0 degrees for $\alpha=0$ (i.e.,
$\langle{\bf S}\rangle || {\bf P}_D$), up to about 6 degrees for $\alpha=40$
along the 2DEG edge. This value might seem rather small for device applications,
however it can be
increased by optimizing the QW configuration. This can be done by taking
advantage of the intrinsic anisotropies of the system.

The presence of the intrinsic bulk inversion asymmetry (BIA), caused by
the underlying crystal structure can be used to enhance the effect of
the Rashba field and increase the amplitude of the precession angle. The
direction of the effective field due to BIA depends on the crystallographic
orientation of the QW.  In a (110)-oriented QW, the BIA effective field is along
the external magnetic field and its direction is antiparallel to B for $k_x=0$
and large positive $k_y$ (i.e., $k_y \parallel [\overline{1}10]$ direction).
In this configuration, its effect is therefore to partially counterbalance
the Zeeman term (i.e., equivalent to increasing the $g$ factor to an effective
value $g^*>g$), making it easier for the spin to bend under the SO
field, and resulting in an increase of the value of $\theta_z$ as shown in
Fig.~\ref{fig:teta}.

Low carrier concentrations and high mobilities have been achieved in
2DEGs in [110]-grown InAs QWs very close to the surface \cite{getzlaff_prb63}.
Using spin-polarized scanning tunneling microscopy \cite{wiebe_rsi75}
on these systems it should be possible to image the spin configurations
shown in Fig.~\ref{fig:spins_xy}-\ref{fig:teta}.

In conclusion we presented a many-body approach to the effects of SO coupling
on the spin configuration in the $\nu=1/3$ FQH state, including electron-electron
interactions. Possible device applications of such a system are suggested
together with experimental techniques to image the calculated spin distribution.

The authors thank D. Grundler, V. Apalkov and P.A. Maksym for their
valuable suggestions and comments.
The work of T.C. has been supported by the Canada
Research Chair Program and the Canadian Foundation for Innovation
Grant. The work of M.C. has been supported by NSERC.


\begin{thebibliography}{99}
\bibitem[\ddag]{byline} Electronic mail:
tapash@physics.umanitoba.ca
\bibitem{awschalom_book}{\em Semiconductor Spintronics and Quantum Computation},
edited by D. D. Awschalom, D. Loss, and N. Samarth (Springer-Verlag, Berlin, 2002).
\bibitem{datta_apl56}S. Datta and B. Das, Appl. Phys. Let. {\bf 56}, 665 (1990).
\bibitem{winkler_prb69}R. Winkler, Phys. Rev. B {\bf 69}, 045317 (2004).
\bibitem{koga_prbR70}T. Koga, J. Nitta, M. van Veerhuizen,
Phys. Rev. B {\bf 70}, 161302R (2004).
\bibitem{book}
T. Chakraborty and P. Pietil\"ainen, {\it The Quantum Hall Effects}
(Second Edition, Springer 1995).
\bibitem{dyakonov_jetp33}M.I. Dyakonov and V.I. Perel, Sov. Phys.
JETP {\bf 33}, 1053 (1971).
\bibitem{recher_PRL85}P. Recher et al. 
Phys. Rev. Lett. {\bf 85}, 1962 (2000).
\bibitem{vandersypen_quantph}L. M. K. Vandersypen et al. 
quant-ph/0207059 (10 Jul. 2002).
\bibitem{ciorga_condmat9912446}M. Ciorga et al.,
cond-mat/9912446 (19 Apr. 2000).
\bibitem{murakami_science301}S. Murakami,
N. Nagaosa and S.-C. Zhang, Science {\bf 301}, 1348 (2003).
\bibitem{pala_PRB71}M. G. Pala,
M. Governale, U. Z\"ulicke and G. Iannaccone,
Phys. Rev. B {\bf 71}, 115306 (2005).
\bibitem{noi_cond-mat}M. Califano, T. Chakraborty and P. Pietil\"ainen,
cond-mat/0411524 (unpublished).
\bibitem{rashba_jpc17}
Y.A. Bychkov and E.I. Rashba, J. Phys. C {\bf 17}, 6039 (1984).
\bibitem{laughlin}
R.B. Laughlin, Phys. Rev. Lett. {\bf 50}, 1395 (1983).
\bibitem{wolf_science294}
S.A. Wolf et al. Science {\bf 294}, 1488 (2001).
\bibitem{getzlaff_prb63}M. Getzlaff {\em et al.},
Phys. Rev. B {\bf 63}, 205305 (2001).
\bibitem{wiebe_rsi75}J. Wiebe, {\em et al.},
Rev. Sci. Instrum. {\bf 75}, 4871 (2004).
\end{thebibliography}
\end{document}